\documentclass[%
 reprint,
 amsmath,amssymb,
 aps,
 showpacs,
 longbibliography,
]{revtex4-1}
\usepackage{graphicx}
\usepackage{natbib}
\usepackage{epsfig}
\usepackage{rotate}
\usepackage{float}
\usepackage{amssymb}
\usepackage{amsbsy}
\usepackage{overpic}
\usepackage{marvosym}
\usepackage{textcomp}
\usepackage{amsmath}
\usepackage{longtable}
\usepackage{graphicx}
\usepackage{dcolumn}
\usepackage{bm}

\newcommand\Nu{\text{Nu}}
\newcommand\Ra{\text{Ra}}
\newcommand\Rey{\text{Re}}
\newcommand\Pran{\text{Pr}}

\usepackage{color}

\begin{document}

\preprint{APS/123-QED}

\title{Mean Temperature Profiles in Turbulent Thermal Convection}%

\author{Olga Shishkina}
 \email{Olga.Shishkina@ds.mpg.de}
\affiliation{Max Planck Institute for Dynamics and Self-Organization,
Am Fassberg 17, 37077 G\"ottingen, Germany}%

\author{Susanne Horn}
 \email{SusanneHorn@ucla.edu}
\affiliation{Earth, Planetary, and Space Sciences, University of California, Los Angeles, USA}%

\author{Mohammad S. Emran}
 \email{Mohammad.Emran@ds.mpg.de}
\affiliation{Max Planck Institute for Dynamics and Self-Organization ,
Am Fassberg 17,  37077 G\"ottingen, Germany}%

\author{Emily S. C. Ching}
 \email{ching@phy.cuhk.edu.hk}
\affiliation{Department of Physics, The Chinese University of Hong Kong, Shatin, Hong Kong}%

\date{\today}

\begin{abstract}
To predict the mean temperature profiles in
turbulent thermal convection, the thermal boundary layer (BL)
equation including the effects of fluctuations has to be solved.
In~[Shishkina {\it et al.}, Phys. Rev. Lett. 114 (2015)], the
thermal BL equation with the fluctuations taken into account as an
eddy thermal diffusivity has been solved for large Prandtl-number
fluids for which the eddy thermal diffusivity and the velocity
field can be approximated respectively as a cubic and a linear
function of the distance from the plate.
In the present work we make use of the idea of Prandtl's mixing
length model and relate the eddy thermal diffusivity to the stream
function. With this proposed relation, we can solve the thermal BL
equation and obtain a closed-form expression for the dimensionless
mean temperature profile in terms of two independent parameters
for fluids with a general Prandtl number. With a proper choice of
the parameters, our predictions of the temperature profiles are in
excellent agreement with the results of our direct numerical
simulations for a wide range of Prandtl numbers 
from 0.01 to 2547.9 and Rayleigh numbers from $10^7$ to $10^9$.

\end{abstract}

\maketitle

\section{Introduction}

Turbulent thermal convection is a major topic in geophysical and
astrophysical fluid dynamics and an important problem in
engineering and technological applications.
The classical systems to study turbulent thermal convection
are Rayleigh--B\'enard convection (RBC) \cite{Ahlers2009,
Bodenschatz2000, Chilla2012, Grossmann2000, Ching2014} where a
fluid is confined between a heated bottom plate and a cooled top
plate and horizontal convection (HC) \cite{Hughes2008,
Shishkina2016a, Shishkina2016} in which the fluid is heated at one
end of the bottom plate and cooled at the other end of the bottom plate.

One important and well-studied question in turbulent thermal
convection research~\cite{Grossmann2000, Grossmann2001,
Stevens2013, Shishkina2016a} is how the mean convective heat and
momentum transport, represented by the Nusselt number ($\Nu$) and
Reynolds number ($\Rey$) respectively, depend on the main input
parameters of the system, which are the Rayleigh number $\Ra\equiv
\alpha g \Delta H^3/(\kappa \nu)$ and the Prandtl number
$\Pran\equiv\nu/\kappa$. Here $\nu$ denotes the kinematic
viscosity, $\kappa$ the thermal diffusivity, $\alpha$ the isobaric
thermal expansion coefficient of the fluid, $g$ the acceleration
due to gravity, $H$ the distance between the heated plate (part)
and the cooled plate (part) for RBC (HC), and $\Delta\equiv
T_h-T_c>0$\, with $T_h$ and $T_c$ respectively the temperatures
of the heated plate (part) and the cooled plate (part) for RBC
(HC). The dependence of $\Nu$ and $\Rey$ on $\Ra$ and $\Pran$ is
influenced significantly by the imposed boundary conditions
\cite{Grossmann2011, Hassanzadeh2014, Gibert2006, Daya2001,
He2011, Boffetta2012, Doering2006, Shishkina2016b,
Shishkina2016c}. Grossmann and Lohse developed a scaling theory
(GL) \cite{Grossmann2000, Grossmann2001} for RBC, which nowadays
allows one to predict $\Nu$ and $\Rey$ if  the pre-factors
\cite{Stevens2013} fitted with the latest experimental and
numerical data are used. The GL theory was later extended to the
case of HC \cite{Shishkina2016a} and magnetoconvection
\cite{Zuerner2016}.

Closely related to the scaling problem of the heat and momentum
transport in different convective systems is the problem to
predict the spatial profiles of the mean flow characteristics.
Among which, the  time- and horizontally area-averaged profile of
the temperature as a function of the vertical distance $z$ from
the heated bottom plate is of particular research interest. In the
Oberbeck--Boussinesq approximation of RBC, the mean temperature
depends only weakly on $z$ in the core part of the domain. There
exists a certain region in the bulk in which
 the mean temperature behaves as
a logarithmic function of $z$ \cite{Ahlers2014}, and this
logarithmic region is expected to  almost fill the entire bulk for
very large $\Ra$ \cite{Grossmann2012}. Near to the bottom and top
plates, the mean temperature changes much more rapidly with $z$
than in the bulk. The knowledge of these boundary layer (BL)
profiles of the mean temperature near the bottom and top plates is
important for many engineering applications as well as for the
development of reliable turbulence models for thermal convection.
It remains one of the most challenging unsolved problems to
predict the mean temperature boundary layer profiles.

In its derivation of the heat and momentum transport scalings in
BL-dominated regimes in RBC, the GL theory \cite{Grossmann2000,
Grossmann2001} assumes that the viscous BL thickness is
proportional to $\Rey^{-1/2}$. This scaling relation
holds, in particular, in the classical
Prandtl-Blasius (PB) boundary-layer theory \cite{Prandtl1905, Landau1987} for steady flows.
The mean temperature profiles obtained in experimental and numerical studies have been compared against the profiles obtained from the PB theory, and
systematic deviations  were reported
\cite{Shishkina2009, Shi2012, Scheel2012,
Stevens2012, Kaczorowski2011, Ovsyannikov2016}. The deviations are generally larger
for larger $\Ra$ and smaller $\Pran$ and they remain even after an
application of a dynamical rescaling procedure \cite{Zhou2010}
that takes into account the time variation of the BL thickness.

In the PB theory, the pressure gradient vanishes and fluctuations do not exist.
The effect of a non-zero pressure gradient within the BLs, or
equivalently the effect of a large-scale mean circulating flow
that is not parallel to the isothermal plate, was studied in
\citet{Shishkina2013} and led to the BL equations, which are
similar to those of Falkner and Skan (FS) \cite{Falkner1931}. With
the FS approach one calculates the ratio of the thermal to viscous
BL thicknesses more accurately compared to PB but the FS
approximation does not lead to a significantly better prediction
of the mean temperature profiles, as the limits of the PB and FS
profiles for infinitesimal $\Pran$ are the same
\cite{Shishkina2014}.
For large $\Pran$ and a flow with a constant shear rate,
\citet{Shraiman1990} derived the mean temperature profile and the
relation between the heat flux and shear rate in thermal
convection. Their mean temperature profile also coincides with the
PB prediction for infinitely large $\Pran$ \cite{Shishkina2009}. \citet{Ching1997}
generalized their approach to the case of a position-dependent
shear rate and derived the temperature profile as a function of
two parameters which are associated with the local thermal BL
thickness and the shear rate. Good agreement of these derived
profiles with the actual ones can be obtained only when the two
parameters are taken as free fitting parameters.

In \citet{Shishkina2015} we derived a new thermal BL equation for
turbulent RBC in large-$\Pran$ fluids. The equation takes into
account the effect of fluctuations, which are neglected in the PB
or FS BL equations, using an eddy thermal diffusivity. In the case
of large $\Pran$, the thermal BL is nested within the viscous BL, thus the eddy thermal
diffusivity and the horizontal mean velocity can be approximated
respectively as a cubic and a linear function of the distance from
the plate.
For the limits $\Pran\gtrsim 1$ and $\Pran\rightarrow\infty$ of
such simplification of the BL equations for large $\Pran$, the
mean temperature profiles were analytically obtained and shown to
be in very good agreement with the profiles obtained in Direct
Numerical Simulations (DNS) of RBC for, respectively, $\Pran =
4.38$ (water) and $\Pran = 2547.9$ (glycerol).

In the present paper we shall derive a thermal BL equation for
fluids with a general $\Pran$, including very small $\Pran$. We
extend the approximation of the eddy thermal diffusivity to larger
$z$ and propose an approximate relation between the
eddy thermal diffusivity and the stream function within the
thermal BL. Then we can solve the resulting thermal BL equation to
obtain the mean temperature profiles in terms of two independent
parameters.
With a proper choice of the parameters, our theoretical
predictions are in perfect agreement with the mean temperature
profiles obtained in the DNS for $\Pran$ down to 0.01. Our present
approach can be reduced to that of \cite{Shishkina2015} in the
case of large $\Pran$.

\section{Basic equation}

Following \cite{Shishkina2015}, we consider the quasi
two-dimensional fluid flow along a semi-infinite horizontal heated
plate and assume that far away from the plate, there exists a
constant mean velocity, the wind, along a horizontal $x$-direction
$x$. The equation for the
temperature field $T(x,z,t)$  is
\begin{equation}
\label{energy1}
\partial_t T + {\bf u}\cdot\nabla T = \kappa \nabla^2 T ,
\end{equation}
where ${\bf u}(x,z,t)\equiv u(x,z,t)\,\hat{x} + v(x,z,t)\,\hat{z}$
is the velocity field, and the flow is incompressible:
\begin{equation} \label{continuity} \nabla \cdot {\bf u} = 0 .
\end{equation} Using Reynolds decomposition of the flow fields
into sums of time-averages and fluctuations,
\begin{eqnarray}
\label{ReynoldsDecomposition}
u = U + u', \quad
 v = V + v', \quad
  T = \Theta + \theta',
\end{eqnarray}
in (\ref{energy1}) and averaging it in time afterwards, we obtain
the following equation for the time-averaged temperature:
\begin{eqnarray}
\label{energy2}
U \partial_x \Theta +
 V \partial_z \Theta +
\partial_x \langle u' \theta'
 \rangle_t &+& \partial_z \langle v'  \theta'
\rangle_t \nonumber\\&=& \kappa  \partial^2_z \Theta+ \kappa  \partial^2_x \Theta,
\end{eqnarray}
where $\langle\cdot\rangle_t$ denotes the time-averaging. The
continuity equation (\ref{continuity}) holds for both the mean and fluctuating velocities:
\begin{eqnarray}
\label{continuity1}
\partial_xU+\partial_zV &=&0, \\
\label{continuity2}
\partial_xu'+\partial_zv'&=&0.
\end{eqnarray}
As usual for BLs,
we assume that within the BL $|\partial_x^2 \Theta| \ll
|\partial_z^2 \Theta|$ and $|\partial_x\langle u'
\theta'\rangle_t|\ll|\partial_z \langle v' \theta' \rangle_t|$ and
obtain
\begin{eqnarray}
\label{energy3}
U \partial_x \Theta +
 V \partial_z \Theta +
 \partial_z \langle v'  \theta'
\rangle_t &=& \kappa  \partial^2_z \Theta.
\end{eqnarray}
Introducing the eddy thermal diffusivity $\kappa_t(x,z)$ for the
fluctuation term in (\ref{energy3}), which is defined by
\begin{eqnarray}
\label{eddykappa}
\langle v' \theta' \rangle_{t} &\equiv& -\kappa_{t} \partial_z \Theta,
\end{eqnarray}
we obtain
\begin{eqnarray}
\label{energy4}
U \partial_x \Theta+ (V-\partial_z \kappa_{t})
\partial_z \Theta &=& (\kappa+\kappa_{t}) \partial_z^2 \Theta.
\end{eqnarray}
In the BL in turbulent thermal convection
the eddy thermal diffusivity is not
negligible~\cite{Shishkina2015}. To satisfy (\ref{continuity1}) we
introduce the stream function $\Psi$, such that
\begin{equation}
 \label{uv}
 U=\partial_z\Psi,\quad V=-\partial_x\Psi .
\end{equation}
We define the
 similarity variable $\xi$ and the dimensionless
stream function $\psi(\xi)$ and temperature
$\theta(\xi)$: \begin{eqnarray}
 \label{xi}
 \xi&\equiv&z/\lambda(x),\\
 \label{psi}
 \Psi&\equiv&U_0 \lambda(x) \psi(\xi),\\
 \label{theta}
 \Theta&\equiv&T_{h}-(\Delta/2) \theta(\xi).
\end{eqnarray}
and look for a similarity solution of (\ref{energy4}) in terms of  $\xi$. Here $\lambda(x)$ is the local thickness of the thermal BL, $U_0$
is the maximal horizontal velocity (wind velocity), $T_h$ is the
temperature of the heated bottom plate, and $\Delta/2$ is the
temperature difference between the bottom plate and the bulk of
the flow. Substituting (\ref{xi})--(\ref{theta}) into
(\ref{energy4}), we obtain the following dimensionless thermal BL equation:
\begin{eqnarray}
 \label{energy5}
(1+\kappa_{t}/\kappa)\theta_{\xi\xi}+[{(\kappa_t/\kappa)_\xi}+B\psi]\theta_\xi&=&0
\end{eqnarray}
with
\begin{equation}
\label{B1}
B={U_0\lambda\lambda_x}/{\kappa}.
\end{equation}
The subscripts $\xi$ and $x$ denote the ordinary derivative with respect to $\xi$ and $x$.
For a similarity solution to exist, $\kappa_t/\kappa$ should
depend on $\xi$ only and $B$ must be a constant, independent of
$x$, therefore $\lambda(x) \propto \sqrt{x}$. With
\begin{equation}
\lambda(x) \propto\sqrt{{\nu x}/{U_0}} \label{lambda},
\end{equation}
from (\ref{B1}) we obtain that $B\propto\Pran$.
It follows from (\ref{lambda}) that the viscous BL thickness scales
as $\Rey^{-1/2}$, where $\Rey \equiv U_0 x/\nu$, if the ratio of
thicknesses of the viscous and thermal BLs depends only on $\Pran$.

To solve the thermal BL equation (\ref{energy5}) with the boundary
conditions
\begin{eqnarray}
\theta(0) = 0, \quad \theta_\xi(0) = 1, \quad \theta(\infty) = 1 ,
\label{BCs}
\end{eqnarray} and obtain the dimensionless temperature profiles $\theta(\xi)$, we need to
know $\kappa_{t}(\xi)/\kappa$ and
$\psi(\xi)$.
In the next two sections we will establish an approximation of
$\kappa_{t}(\xi)$ and propose an approximate relation between $\kappa_t(\xi)$
and $\psi(\xi)$.

\section{Eddy thermal diffusivity}

Very close to the plate, the eddy thermal diffusivity can be
approximated as a cubic function of $\xi$ \cite{Shishkina2015}.
In this regard, the eddy thermal diffusivity and the eddy viscosity
exhibit similar behavior near the plate \cite{Antonia1991}. From the continuity
equation~(\ref{continuity2}) of the fluctuating velocity
it follows that $\partial_zv'=0$ at the plate ($z=0$). From this
result and the fact that all fluctuations $u'$, $v'$ and $\theta'$
vanish at $z=0$, we obtain consequentially
\begin{eqnarray}
\label{qqq4}
 \langle v' \theta' \rangle_{t}\big|_{z=0}=0, \
\label{qqq5}
 \partial_z\langle v' \theta' \rangle_{t}\big|_{z=0}=0,\
\label{qqq6}
 \partial_z^2\langle v' \theta' \rangle_{t}\big|_{z=0}=0. \qquad
  \end{eqnarray}
Using the definition of $\kappa_t$ (\ref{eddykappa}) and the
linear dependency of $\xi$ on $z$~[see (\ref{xi})], these results imply
\begin{eqnarray}
\kappa_t\big|_{\xi=0}=\left(\kappa_t\right)_\xi\big|_{\xi=0}=\left(\kappa_t\right)_{\xi\xi}\big|_{\xi=0}=0
\qquad
\end{eqnarray}
and, hence, close to the plate, $\kappa_t/\kappa$ can be approximated as a cubic function of $\xi$,
\begin{eqnarray}
\label{bigOkappa222}
\kappa_t/\kappa\approx a^3\xi^3,
\end{eqnarray}
with a certain constant $a$, which measures the size of fluctuations.

Relatively far away from the plate, the mean temperature
$\Theta$ behaves as a logarithmic function of the distance $z$
from the plate \cite{Grossmann2012, Ahlers2012, He2014}. In this
logarithmic or inner region, the fluctuations are so strong that the term
$\partial_z\langle v' \theta'\rangle_t$ dominates the other terms
on the left-hand side of (\ref{energy3}),
\begin{eqnarray}
\label{energy33}
 \partial_z \langle v'  \theta'
\rangle_t \approx \kappa  \partial^2_z \Theta,
\end{eqnarray}
which implies
\begin{eqnarray}
\label{energy333} \langle v'  \theta' \rangle_t \approx \kappa [\partial_z \Theta -
\left.\partial_z \Theta\right|_{z=0}] \approx - \kappa \left.\partial_z \Theta\right|_{z=0}
\end{eqnarray}
as in the inner region the mean temperature $\Theta$ changes
very slowly with $z$ so that $|\left.\partial_z
\Theta\right|_{z}|\ll|\left.\partial_z \Theta\right|_{z=0}|$.
Using (\ref{eddykappa}) and (\ref{energy333}), we have
\begin{equation}
(\kappa_t/\kappa) \  \partial_z \Theta \approx \partial_z \Theta \big |_{z=0}
\end{equation}
The logarithmic dependence on $z$ of $\Theta$ thus implies that
$\kappa_t/\kappa$ behaves as a linear function of $z$ or $\xi$ in this region:
\begin{equation}
\label{bigOkappa2222} \kappa_t/\kappa \sim \xi.
\end{equation}
These two different behaviors of $\kappa_t/\kappa$ on $\xi$,
(\ref{bigOkappa222}) for small $\xi$ and (\ref{bigOkappa2222}) for
large $\xi$, have both been demonstrated in \cite{Shishkina2015}.

Based on these two behaviors, we make the following
approximation of $(\kappa_t/\kappa)_\xi$:
\begin{eqnarray}
 \label{appr1}
(\kappa_t/\kappa)_\xi \approx\frac{3a^3\xi^2}{1+b^2\xi^2},
\end{eqnarray}
where $b$ is a constant that determines the location $\xi_{\max}$
of the maximum value of $(\kappa_t/\kappa)_{\xi\xi}$, namely
\begin{eqnarray}
 \label{ximax}
\xi_{\max}=(\sqrt{3}b)^{-1}.
\end{eqnarray}
From
(\ref{appr1})
 we obtain
\begin{eqnarray}
 \label{appr2}
\frac{\kappa_t}{\kappa}\approx
\frac{3a^3}{b^3}[b\,\xi-\arctan(b\,\xi)].
\end{eqnarray}
which gives the two limiting behaviors discussed above, namely
\begin{eqnarray}
 \label{appr3}
\frac{\kappa_t}{\kappa}\approx
3a^3\left(
\frac{\xi^3}{3}+
\frac{b^2\xi^5}{5}+
\frac{b^4\xi^7}{7}+\mathcal{O}(\xi^9)\right)\approx a^3\xi^3
\end{eqnarray}
for $\xi\rightarrow0$, and
\begin{eqnarray}
 \label{appr4}
\frac{\kappa_t}{\kappa}\approx
3a^3\left(
\frac{\xi}{b^2}-
\frac{\pi}{2b^3}+
\frac{1}{b^4\xi}+\mathcal{O}(\xi^{-3})\right)\approx \frac{a^3}{b^2}\xi
\end{eqnarray} for $\xi\rightarrow\infty$.

\section{Proposed relation based on mixing length model}

We first make use of the idea of Prandtl's mixing
length model~\cite{Prandtl1925} to relate $\kappa_t$ to the mean
velocity gradient. According to Prandtl's mixing length model, a
fluid parcel will retain its velocity for a mixing length $l_v$
before mixing with surrounding fluid in a turbulent environment.
Thus, the fluctuation in the velocity can be seen as the difference in
velocity between a distance $l_v$. As all fluctuations in the
thermal BL are mostly along the vertical $z$-direction, we use
this picture to approximate the vertical velocity fluctuation $v'$ by
\begin{equation} v' \approx l_v
\partial_z V.
\label{mixing1}
\end{equation}
Similarly, we approximate
the temperature fluctuation $\theta'$ by
\begin{equation}
\theta' \approx - l_\theta \partial_z \Theta, \label{mixing2}
\end{equation}
where $l_\theta$ is the mixing length for temperature.
 Using
(\ref{mixing1}) and (\ref{mixing2}), we have
\begin{equation}
\langle v' \theta' \rangle_t \approx - l_v l_\theta \;\partial_z V \,
\partial_z \Theta. \label{mixing3}
\end{equation}
Comparing (\ref{mixing3}) with (\ref{eddykappa}), we obtain
\begin{equation}
\kappa_t/\kappa \approx (l_v l_\theta/\kappa) \,\partial_z V,
\label{mixing4}
\end{equation}
which relates the eddy thermal diffusivity to the mean velocity
gradient.

Next, we evaluate (\ref{mixing4}) near the plate to get a direct
relation between $(\kappa_t/\kappa)_\xi$ and $\psi$.  Near the
plate, we estimate the mixing lengths to be proportional to $z$:
\begin{equation} \label{approx}
 l_v \approx k_v z , \qquad l_\theta \approx
k_\theta z,
\end{equation}
where $k_v$ and $k_\theta$ are some positive constants.
Substituting (\ref{approx}) into (\ref{mixing4}) and using
(\ref{uv})-(\ref{psi}) and (\ref{B1}), we thus have
\begin{equation} \kappa_t/\kappa \approx B \,k_v k_\theta \,\xi^3\,\psi_{\xi\xi}
\label{mixing5}
\end{equation}
Taking the derivative of (\ref{mixing5}) w.r.t. $\xi$ and keeping only the lowest order term in $\xi$, we obtain
\begin{eqnarray} (\kappa_t/\kappa)_\xi
\approx 3 B \,k_v k_\theta \,\xi^2\,\psi_{\xi\xi}(0)  \qquad
\mbox{for small } \xi. \label{kappaApp}
\end{eqnarray}
On the other hand, using $\psi(0)=\psi_\xi(0)=0$ that result from
the no-slip boundary condition, we obtain
\begin{equation}
\psi \approx \psi_{\xi \xi}(0) \,\xi^2/2  \qquad \mbox{for small } \xi. \label{psiApp}
\end{equation}
Hence, (\ref{kappaApp}) and (\ref{psiApp}) give
\begin{equation}
(\kappa_t/\kappa)_\xi  \approx 6 B \,k_v k_\theta \,\psi \qquad
\mbox{for small }  \xi, \label{relation0}
\end{equation}
establishing a similarity between the dimensionless stream
function $\psi$ and the derivative of the eddy thermal diffusivity
$(\kappa_t/\kappa)_\xi$ near the plate.

The mean horizontal velocity $U$ grows linearly with distance
close to the plate. At a certain distance from the plate it
attains a maximum value (which gives the wind velocity) and then
decays to zero towards the bulk of the flow. From (\ref{uv}) and
(\ref{psi}), $U=U_0\psi_\xi$, and $\xi$ is linearly related with
the vertical coordinate $z$, therefore, the dimensionless stream
function $\psi$ goes as $\xi^2$ near the plate and is almost
constant far away from the plate.  Thus, the functional
dependences of $\psi$ and $(\kappa_t/\kappa)_\xi$ on $\xi$ are similar in two
limits: both of them $\sim\xi^2$ for $\xi\rightarrow0$ and $\sim
const$ for $\xi\rightarrow\infty$. This observation
together with (\ref{relation0}) motivate us to propose the
following approximate relation for the whole thermal BL:
\begin{equation}
(\kappa_t/\kappa)_\xi \approx K B \psi \label{relation}
\end{equation}
for some constant $K > 0$.

\section{Theoretical model}

Using the proposed relation (\ref{relation}), we obtain
\begin{equation}
(\kappa_t/\kappa)_\xi+B\psi \approx (1+1/K)
(\kappa_t/\kappa)_\xi \equiv c (\kappa_t/\kappa)_\xi,
\label{appr0}
\end{equation} and (\ref{energy5}) becomes
\begin{equation}
 \label{energy6}
(1+\kappa_{t}/\kappa)\theta_{\xi\xi}+c(\kappa_t/\kappa)_\xi
\theta_\xi =0.
\end{equation}
Equation (\ref{energy6}) is a thermal BL equation for all values
of $\Pran$, including $\Pran < 1$. When the fluctuations are
relatively weak so that the flow remains in the transition from
laminar to turbulent state, the value of $c$ can be large. When
the fluctuations are so strong that the term
$(\kappa_t/\kappa)_\xi$ dominates $B\psi$ in (\ref{appr0}), the
constant $c$ is close to 1. The solution of (\ref{energy6}) is
\begin{eqnarray}
 \label{solution}
\theta(\xi)=
\int_0^\xi\,\Bigl[1+\frac{\kappa_t}{\kappa}(\eta)\Bigr]^{-c}\,d\eta,
\end{eqnarray}
which together with the approximation (\ref{appr2}) yields
\begin{eqnarray}
\theta(\xi)
 \label{solution22}
&=&\frac{1}{b}\int_0^{b\,\xi}\,\Bigl[1+\frac{3a^3}{b^3}(\eta-\arctan(\eta))\Bigr]^{-c}\,d\eta.
\end{eqnarray}
Note that (\ref{solution22}) has two independent parameters only
as the following must be fulfilled,
\begin{eqnarray}
 \label{solution3}
b&=&\int_0^\infty\,\Bigl[1+\frac{3a^3}{b^3}(\eta-\arctan(\eta))\Bigr]^{-c}\,d\eta,
\end{eqnarray}
due to the boundary conditions far away from the plate,
$\theta(\infty)=1$.

For the particular case of very large $\Pran$, the thermal BL is
deeply nested within the viscous BL and the eddy thermal
diffusivity can be approximated by (\ref{appr3}). Thus,
(\ref{solution}) is reduced to the form reported in
\cite{Shishkina2015}:
\begin{equation}
\label{PRL} \theta(\xi)=\int_0^\xi (1+a^3\eta^3)^{-c}d\eta,
\end{equation}
where the constants $a$ and $c$ are related with
\begin{eqnarray}
\label{a}
a=\frac{\Gamma\left({1}/{3}\right)\Gamma\left(c-{1}/{3}\right)}{3\Gamma\left(c\right)}
\end{eqnarray}
and $\Gamma$ is the gamma function. As discussed in
\cite{Shishkina2015}, analytical expressions for $\theta$ can be
obtained from~(\ref{PRL}) for $c=1$, which corresponds to the
limiting case of large fluctuations:
\begin{eqnarray}
\label{c1}
\theta&=&\frac{\sqrt{3}}{4\pi}
\log\frac{(1+e\xi)^3}{1+(e\xi)^3}+ \frac{3}{2\pi}\arctan
\frac{2e\xi-1}{\sqrt{3}}
+\frac{1}{4}\qquad
\end{eqnarray}
with $e={2\pi}/({3\sqrt{3}})\approx1.2$ as well as for $c=2$:
\begin{eqnarray}
\nonumber  \theta=\frac{\sqrt{3}}{4\pi}
\log\frac{(1+f\xi)^3}{1+(f\xi)^3}+ \frac{3}{2\pi}\arctan
\frac{2f\xi-1}{\sqrt{3}}\quad
\\
+ \frac{\xi}{3(1+(f\xi)^3)}+\frac{1}{4} \label{c2}
\end{eqnarray}
with $f={4\pi}/({9\sqrt{3}})\approx 0.8$.

Equations (\ref{c1}) and (\ref{c2}) are found to be in good
agreement with DNS results for $\Pran=4.38$ and $\Pran=2547.9$
respectively, as reported in \cite{Shishkina2015}.
For intermediate values of $\Pran$ between 4.38 and 2547.9,
(\ref{PRL}) with a fitted value of $c$ is shown to be in good
agreement with DNS results~\cite{Ching2017}.

\section{Validation of the model}

In \cite{Shishkina2015, Ching2017} we have shown that
(\ref{PRL}) describes the temperature profiles obtained in DNS of
RBC very well in the large-$\Pran$ regime, from $\Pran=4.38$ to
$\Pran=2547.9$. The DNS simulations were conducted in a
cylindrical container
with a diameter-to-height aspect ratio 1,
using the finite-volume
computational code \textsc{Goldfish} \cite{Kooij2017}. This code
features a high flexibility in the choice of the size of the
computational grids, which are finer near the domain boundaries and resolve
the Kolmogorov and Batchelor microscales
\cite{Shishkina2010}. Our present work aims for a prediction of
the temperature profiles for general $\Pran$, and in particular
for small $\Pran$. For this purpose, additional simulations for
$\Pran=0.01$, $\Pran=0.0232$, $\Pran=0.1$ and $\Pran=1$ have been
performed. Here we will check the new 
results (\ref{solution22}), (\ref{solution3})
mostly against these additional DNS data of small $\Pran$, 
and consider only two cases of large $\Pran$
(see Table~\ref{table1} for the details of the cases studied).
 We will also show that for large $\Pran$ the
new profiles (\ref{solution22}) with (\ref{solution3}) are very
close to our earlier result (\ref{PRL}) reported in
\cite{Shishkina2015}.

\begin{figure}
\centering \includegraphics[width=0.45\textwidth]{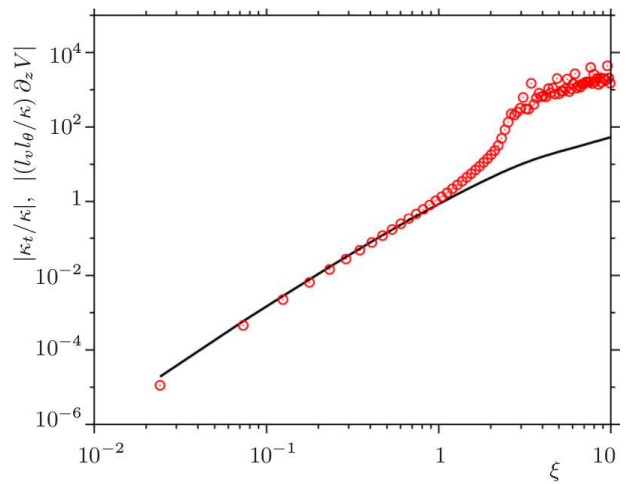}
\caption{Normalized eddy thermal diffusivity
$|\kappa_{t}/\kappa|$, calculated for $\kappa_{t} =
\left(V\Theta-\langle v T\rangle_{t}\right)/\partial_z \Theta$,
and then averaged over horizontal cross-sections, obtained in the
DNS for $\Pran=10$ and $\Ra=10^8$ (symbols) together with a fit
for $|(l_v l_\theta/\kappa) \,\partial_z V|$ averaged over
horizontal cross-sections (solid line). Here $l_v$ and $l_\theta$
are taken according to (\ref{approx}), i.e. $l_vl_\theta\sim z^2$. Thus, the symbols and the line represent,
respectively,  the magnitudes of the left- and right-hand sides of the assumption
(\ref{mixing4}). } \label{PIC1a}
\end{figure}

We first check directly the validity of
(\ref{mixing4}) with $l_v$ and $l_\theta$ given by the
approximation (\ref{approx}), which form the basis of our proposed
relation~(\ref{relation}). As shown in Fig.~\ref{PIC1a},
(\ref{mixing4}), with $l_v$ and $l_\theta$ given by (\ref{approx})
with $k_vk_\theta\approx1$ indeed holds well near the
plates up to $\xi \approx 1$.

\begin{figure}
\centering \includegraphics[width=0.45\textwidth]{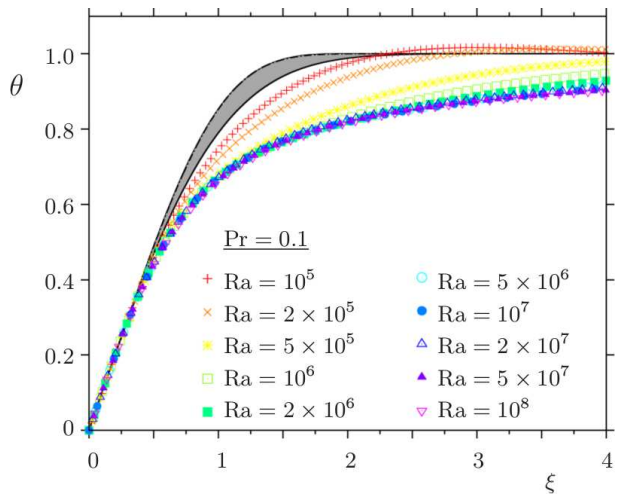}
\caption{Temperature profiles, averaged in time and over horizontal cross sections,
obtained in the DNS of RBC in a cylindrical container of the aspect ratio 1 for $\Pran=0.1$ and different $\Ra$ (symbols).
One can see that the profiles converge with increasing $ \Ra$.
Prandtl--Blasius predictions for $\Pran\rightarrow\infty$ ($-\cdot-$) and $\Pran\rightarrow0$ (---)
bound the gray region of Prandtl--Blasius predictions for all intermediate $\Pran$.
}
\label{PIC1}
\end{figure}

Then we check our new results (\ref{solution22}),
(\ref{solution3}) against the DNS data.
We normalize the mean temperature profiles
$\theta$, obtained in the DNS and averaged in time and over
horizontal cross sections, in such a way that $\theta$ is equal to
0 at the plate and to 1 in the central part of the domain and its
derivative with respect to $\xi$ in the vertical direction is
equal to 1 at the plate. For a fixed $\Pran$ and varying $\Ra$,
the profiles are generally different, see Fig.~\ref{PIC1}. In
laminar and transitional regimes, the mixing in the core part of
the domain is limited and different complicated global flow
structures develop, which for smaller $\Ra$ in some cases may even
cause overshoot profiles. With increasing $\Ra$, the temperature
profiles start to converge. Thus, in turbulent regime, starting at
a certain sufficiently large $\Ra$, all temperature profiles
almost coincide. In Fig.~\ref{PIC1} one can see that for
$\Pran=0.1$ the temperature profiles differ significantly for
$\Ra$ from $10^5$ to $10^6$ and almost replicate each other for
$\Ra\geq5\times10^6$. All of them lie outside the region of the
Prandtl--Blasius (or Falkner--Skan) predictions for all possible
$\Pran$, which is shown in gray colour in Fig.~\ref{PIC1}. For
smaller $\Ra$, the profiles are closer to the PB predictions but
the converged temperature profiles clearly lie far outside of the
PB region.

\begin{figure}
\centering \includegraphics[width=0.45\textwidth]{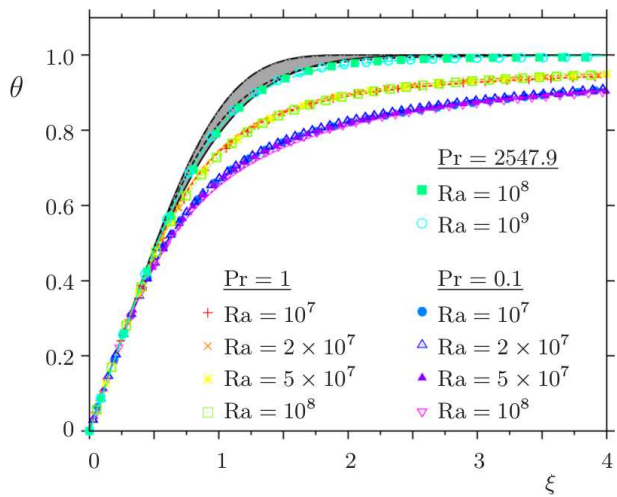}
\caption{Temperature profiles, averaged in time and over
horizontal cross sections, obtained in the DNS of RBC in a
cylindrical container of the aspect ratio 1 for $\Pran=0.1$,
$\Pran=1$ and $\Pran=2547.9$ and different $\Ra$ (symbols)
together with the predictions (\ref{solution22}) (lines of the
corresponding colours), see Table~\ref{table1}. The black dashed
line corresponds to the simplification (\ref{c2}) for very large
$\Pran$, as reported in \cite{Shishkina2015}. The Prandtl--Blasius
region (gray) is as in Fig.~\ref{PIC1}. }
\label{PIC2}
\end{figure}

We focus now on the temperature profiles of the fully developed
turbulent convective flows, i.e. on the converged profiles for
sufficiently large $\Ra$. As discussed, these profiles depend
strongly on $\Pran$ with little or no dependence on $\Ra$. In
Fig.~\ref{PIC2} such profiles are presented for $\Pran=0.1$,
$\Pran=1$ and $\Pran=2547.9$, as obtained in the DNS (symbols)
together with the model solutions of (\ref{solution22}) (lines of
the corresponding colours) for proper choices of the parameters
$a$, $b$ and $c$. More precisely, the parameter $a$ is found by
fitting the DNS data for $\kappa_t/\kappa$ by the approximation
(\ref{appr3}) near the heated plate within half the thermal BL
thickness, i.e. for $\xi$ ranging from 0 to 0.5. Then the
parameters $b$ and $c$ are sought by fitting (\ref{solution22}) to
the DNS temperature profiles, while varying $c$ and keeping $b$
satisfying the equation (\ref{solution3}). The fitted values of
$a$, $b$ and $c$ obtained for different $\Ra$ and $\Pran$ are
presented in Table~\ref{table1}.
Evidently, (\ref{solution22}) perfectly describes
the temperature profiles in a wide range of $\Pran$ including $\Pran\ll1$.

\begin{table}[h!]
\begin{tabular}{ccccc}
 $\qquad\Pran\qquad$&$\qquad\Ra\qquad$&$\qquad a\qquad$&$\qquad b\qquad$&$\qquad c\qquad$\\
 \hline\\
 0.01 & $10^7$ & 1.59 & 6.19 & 4.99\\
 0.0232 & $10^7$ & 1.56 &3.59 & 2.64\\
 0.1 & $10^7$ & 1.52 & 2.27 & 1.84\\
 0.1 & $2\times10^7$ & 1.49 & 2.21 & 1.86\\
 0.1 & $5\times10^7$ & 1.59 & 2.72 & 1.97\\
 0.1 & $10^8$ & 1.62 & 2.79 & 1.96\\
 1 & $10^7$ & 1.16  & 0.62 & 1.36\\
 1 & $2\times10^7$ & 1.13 & 0.63 & 1.41\\
 1 & $5\times10^7$ & 1.12 & 0.81& 1.57\\
 1 & $10^8$ & 1.15 & 0.64 & 1.39\\
 4.38 & $10^8$ & 1.00 & 0.61 & 1.68  \\
 4.38 & $10^9$ & 1.02 & 0.62 & 1.64  \\
 2547.9 & $10^8$ & 0.77 & 0.51 & 2.61\\
 2547.9 & $10^9$ & 0.75 & 0.52 & 2.77\\
 \hline
\end{tabular}
\caption{
Fitted values of the parameters in the
temperature profiles approximation (\ref{solution22}).
\label{table1} }
\end{table}

\begin{figure}
\centering \includegraphics[width=0.45\textwidth]{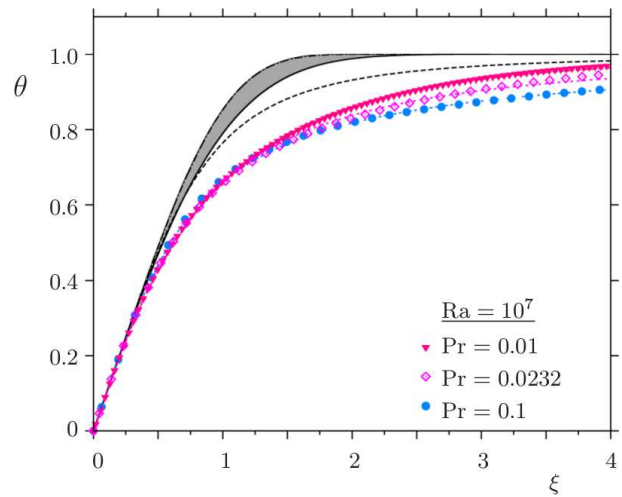}
\caption{Temperature profiles, averaged in time and over horizontal cross sections,
obtained in the DNS of RBC in a cylindrical container of the aspect ratio 1 for $\Ra=10^7$ and different small Prandtl numbers (symbols)
together with the predictions (\ref{solution22}) from the present work (lines of the corresponding colours), see Table~\ref{table1}.
The black dashed line corresponds to the limiting case (\ref{c1}) for the simplification (\ref{PRL}) for large $\Pran$.
The Prandtl--Blasius region (gray) is as in Fig.~\ref{PIC1}.
}
\label{PIC3}
\end{figure}

Finally, we consider transitional RBC flows with very small
$\Pran$, specifically $\Pran=0.01$ and $\Pran=0.0232$ for $\Ra=10^7$.
Also for these RBC flows, the temperature profiles are in excellent agreement with \eqref{solution22},
as illustrated in Fig.~\ref{PIC3}.
The corresponding parameters are given in Table~\ref{table1}.

\section{Conclusions}
Utilizing the idea of Prandtl's mixing length model, we put forward
an approximate relation between the eddy
thermal diffusivity $\kappa_t$ and the stream function $\psi$
within the thermal BL. This proposed relation has allowed
us to obtain a thermal BL equation (\ref{energy6}) that takes
fluctuations for fluids with a general $\Pran$ into account, thus extending our
earlier work \cite{Shishkina2015}.
Using the present approximation (\ref{appr2}) for the eddy thermal
diffusivity for the entire thermal BL, we have obtained the
solution (\ref{solution22}) of the thermal BL equation
(\ref{energy6}) in terms of two independent parameters $a$ and
$c$.  The third parameter, $b$, is fixed by the boundary condition
far away from the plate, which is given by (\ref{solution3}). The
parameter $a$ measures  the intensity of the fluctuations very
close to the plate according to (\ref{appr3}) while the parameter
$c\geq1$ reflects the relative magnitudes of the stream function
and the derivative of the eddy thermal diffusivity. When the BL
flow is highly fluctuating so that the term
$(\kappa_t/{\kappa})_\xi$ dominates the term $B\psi$ in
 (\ref{appr0}), $c$ is about 1. On the other hand, in a
transitional flow with relatively weak fluctuations, the value of
$c$ is large. With a proper choice of $a$ and $c$, our theoretical
model (\ref{solution22}) describes extremely precisely the
temperature profiles near the heated or cooled horizontal plates,
in transitional and turbulent convective flows, for very large as
well as very small $\Pran$.
In the present work, we have obtained the fitted value of $a$ by using DNS data of
$\kappa_t/\kappa$ near the heated plate. In situations where
measurements of $\kappa_t$ are not available,
as in most experimental studies, we suggest to fit the measured temperature
profiles directly by (\ref{solution22}) with the constraint
(\ref{solution3}) to get the values of the two independent
parameters $a$ and $c$.
Note that our earlier model (\ref{PRL}), (\ref{a}) for the temperature profiles in large-Prandtl-number RBC,
which was proposed in \cite{Shishkina2015}, has only one free parameter, while
the new model (\ref{solution22}), (\ref{solution3}) has two free parameters but is applicable to general $\Pran$.

 To derive empirical formulas for the parameters $a$ and $c$
in the model (\ref{solution22}), (\ref{solution3}),
further experimental and numerical data are needed, in
particular for very high $\Ra$ and very small $\Pran$.
It should be noted that the DNS of RBC by large
$\Ra$ and either very large $\Pran$ \cite{Shishkina2015} or very
small $\Pran$ \cite{Scheel2016, Schumacher2016}  require enormous
computational efforts due to the requirement to resolve all
relevant spatial and temporal microscales \cite{Shishkina2010}.
That is, the time stepping must be finer than the time microscale
$\tau=(\nu/\epsilon_u)^{1/2}$ and the spatial stepping must be
smaller than the Kolmogorov microscale
$\eta=(\nu^3/\epsilon_u)^{1/4}$ if $\Pran\leq1$ or smaller than
the Batchelor microscale $\eta_B=(\nu\kappa^2/\epsilon_u)^{1/4}$
if $\Pran>1$. Here $\epsilon_u$ is the mean kinetic dissipation
rate, which in RBC equals
$\epsilon_u=(\nu^3/H^4)(\Nu-1)\Ra^2\Pran^{-2}$. For large $\Pran$
the need to resolve the time microscale is restrictive, while for
small $\Pran$ the very fine meshes in space are needed to resolve
the Kolmogorov spatial microscale.
The general dependence of the
temperature profiles (\ref{solution22}) on $\Pran$, $\Ra$ and the
geometrical characteristics of the convection cell will
therefore  be explored in future when
more experimental and numerical data, in particular for very small
$\Pran$, will be available.

\begin{acknowledgments}
OS, ME and SH  acknowledge the financial support of the Deutsche Forschungsgemeinschaft (DFG) under grants
Sh405/4-2 (Heisenberg fellowship), Sh405/3-2 and Ho 5890/1-1.
The authors thank the Leibniz Supercomputing Centre (LRZ) for providing computing time.
\end{acknowledgments}

%


\begin{thebibliography}{46}%
\makeatletter
\providecommand \@ifxundefined [1]{%
 \@ifx{#1\undefined}
}%
\providecommand \@ifnum [1]{%
 \ifnum #1\expandafter \@firstoftwo
 \else \expandafter \@secondoftwo
 \fi
}%
\providecommand \@ifx [1]{%
 \ifx #1\expandafter \@firstoftwo
 \else \expandafter \@secondoftwo
 \fi
}%
\providecommand \natexlab [1]{#1}%
\providecommand \enquote  [1]{``#1''}%
\providecommand \bibnamefont  [1]{#1}%
\providecommand \bibfnamefont [1]{#1}%
\providecommand \citenamefont [1]{#1}%
\providecommand \href@noop [0]{\@secondoftwo}%
\providecommand \href [0]{\begingroup \@sanitize@url \@href}%
\providecommand \@href[1]{\@@startlink{#1}\@@href}%
\providecommand \@@href[1]{\endgroup#1\@@endlink}%
\providecommand \@sanitize@url [0]{\catcode `\\12\catcode `\$12\catcode
  `\&12\catcode `\#12\catcode `\^12\catcode `\_12\catcode `\%12\relax}%
\providecommand \@@startlink[1]{}%
\providecommand \@@endlink[0]{}%
\providecommand \url  [0]{\begingroup\@sanitize@url \@url }%
\providecommand \@url [1]{\endgroup\@href {#1}{\urlprefix }}%
\providecommand \urlprefix  [0]{URL }%
\providecommand \Eprint [0]{\href }%
\providecommand \doibase [0]{http://dx.doi.org/}%
\providecommand \selectlanguage [0]{\@gobble}%
\providecommand \bibinfo  [0]{\@secondoftwo}%
\providecommand \bibfield  [0]{\@secondoftwo}%
\providecommand \translation [1]{[#1]}%
\providecommand \BibitemOpen [0]{}%
\providecommand \bibitemStop [0]{}%
\providecommand \bibitemNoStop [0]{.\EOS\space}%
\providecommand \EOS [0]{\spacefactor3000\relax}%
\providecommand \BibitemShut  [1]{\csname bibitem#1\endcsname}%
\let\auto@bib@innerbib\@empty
\bibitem [{\citenamefont {Ahlers}\ \emph {et~al.}(2009)\citenamefont {Ahlers},
  \citenamefont {Grossmann},\ and\ \citenamefont {Lohse}}]{Ahlers2009}%
  \BibitemOpen
  \bibfield  {author} {\bibinfo {author} {\bibfnamefont {G.}~\bibnamefont
  {Ahlers}}, \bibinfo {author} {\bibfnamefont {S.}~\bibnamefont {Grossmann}}, \
  and\ \bibinfo {author} {\bibfnamefont {D.}~\bibnamefont {Lohse}},\ }\bibfield
   {title} {\enquote {\bibinfo {title} {Heat transfer and large scale dynamics
  in turbulent {R}ayleigh--{B}\'{e}nard convection},}\ }\href@noop {}
  {\bibfield  {journal} {\bibinfo  {journal} {Rev. Mod. Phys.}\ }\textbf
  {\bibinfo {volume} {81}},\ \bibinfo {pages} {503--537} (\bibinfo {year}
  {2009})}\BibitemShut {NoStop}%
\bibitem [{\citenamefont {Bodenschatz}\ \emph {et~al.}(2000)\citenamefont
  {Bodenschatz}, \citenamefont {Pesch},\ and\ \citenamefont
  {Ahlers}}]{Bodenschatz2000}%
  \BibitemOpen
  \bibfield  {author} {\bibinfo {author} {\bibfnamefont {E.}~\bibnamefont
  {Bodenschatz}}, \bibinfo {author} {\bibfnamefont {W.}~\bibnamefont {Pesch}},
  \ and\ \bibinfo {author} {\bibfnamefont {G.}~\bibnamefont {Ahlers}},\
  }\bibfield  {title} {\enquote {\bibinfo {title} {Recent developments in
  {R}ayleigh--{B}\'{e}nard convection},}\ }\href@noop {} {\bibfield  {journal}
  {\bibinfo  {journal} {Annu. Rev. Fluid Mech.}\ }\textbf {\bibinfo {volume}
  {32}},\ \bibinfo {pages} {709--778} (\bibinfo {year} {2000})}\BibitemShut
  {NoStop}%
\bibitem [{\citenamefont {Chill\`{a}}\ and\ \citenamefont
  {Schumacher}(2012)}]{Chilla2012}%
  \BibitemOpen
  \bibfield  {author} {\bibinfo {author} {\bibfnamefont {F.}~\bibnamefont
  {Chill\`{a}}}\ and\ \bibinfo {author} {\bibfnamefont {J.}~\bibnamefont
  {Schumacher}},\ }\bibfield  {title} {\enquote {\bibinfo {title} {{New
  perspectives in turbulent Rayleigh--B\'enard convection}},}\ }\href@noop {}
  {\bibfield  {journal} {\bibinfo  {journal} {Eur. Phys. J. E}\ }\textbf
  {\bibinfo {volume} {35}},\ \bibinfo {pages} {58} (\bibinfo {year}
  {2012})}\BibitemShut {NoStop}%
\bibitem [{\citenamefont {Grossmann}\ and\ \citenamefont
  {Lohse}(2000)}]{Grossmann2000}%
  \BibitemOpen
  \bibfield  {author} {\bibinfo {author} {\bibfnamefont {S.}~\bibnamefont
  {Grossmann}}\ and\ \bibinfo {author} {\bibfnamefont {D.}~\bibnamefont
  {Lohse}},\ }\bibfield  {title} {\enquote {\bibinfo {title} {Scaling in
  thermal convection: {A} unifying theory},}\ }\href@noop {} {\bibfield
  {journal} {\bibinfo  {journal} {J. Fluid Mech.}\ }\textbf {\bibinfo {volume}
  {407}},\ \bibinfo {pages} {27--56} (\bibinfo {year} {2000})}\BibitemShut
  {NoStop}%
\bibitem [{\citenamefont {Ching}(2014)}]{Ching2014}%
  \BibitemOpen
  \bibfield  {author} {\bibinfo {author} {\bibfnamefont {E.~S.~C.}\
  \bibnamefont {Ching}},\ }\href@noop {} {\emph {\bibinfo {title} {{Statistics
  and scaling in turbulent Rayleigh--B\'enard convection}}}}\ (\bibinfo
  {publisher} {Springer, Singapore},\ \bibinfo {year} {2014})\BibitemShut
  {NoStop}%
\bibitem [{\citenamefont {Hughes}\ and\ \citenamefont
  {Griffiths}(2008)}]{Hughes2008}%
  \BibitemOpen
  \bibfield  {author} {\bibinfo {author} {\bibfnamefont {G.~O.}\ \bibnamefont
  {Hughes}}\ and\ \bibinfo {author} {\bibfnamefont {R.~W.}\ \bibnamefont
  {Griffiths}},\ }\bibfield  {title} {\enquote {\bibinfo {title} {Horizontal
  convection},}\ }\href@noop {} {\bibfield  {journal} {\bibinfo  {journal}
  {Ann. Rev. Fluid Mech.}\ }\textbf {\bibinfo {volume} {40}},\ \bibinfo {pages}
  {185--208} (\bibinfo {year} {2008})}\BibitemShut {NoStop}%
\bibitem [{\citenamefont {Shishkina}\ \emph {et~al.}(2016)\citenamefont
  {Shishkina}, \citenamefont {Grossmann},\ and\ \citenamefont
  {Lohse}}]{Shishkina2016a}%
  \BibitemOpen
  \bibfield  {author} {\bibinfo {author} {\bibfnamefont {O.}~\bibnamefont
  {Shishkina}}, \bibinfo {author} {\bibfnamefont {S.}~\bibnamefont
  {Grossmann}}, \ and\ \bibinfo {author} {\bibfnamefont {D.}~\bibnamefont
  {Lohse}},\ }\bibfield  {title} {\enquote {\bibinfo {title} {Heat and momentum
  transport scalings in horizontal convection},}\ }\href@noop {} {\bibfield
  {journal} {\bibinfo  {journal} {Geophys. Res. Lett.}\ }\textbf {\bibinfo
  {volume} {43}},\ \bibinfo {pages} {1219--1225} (\bibinfo {year}
  {2016})}\BibitemShut {NoStop}%
\bibitem [{\citenamefont {Shishkina}\ and\ \citenamefont
  {Wagner}(2016)}]{Shishkina2016}%
  \BibitemOpen
  \bibfield  {author} {\bibinfo {author} {\bibfnamefont {O.}~\bibnamefont
  {Shishkina}}\ and\ \bibinfo {author} {\bibfnamefont {S.}~\bibnamefont
  {Wagner}},\ }\bibfield  {title} {\enquote {\bibinfo {title} {Prandtl-number
  dependence of heat transport in laminar horizontal convection},}\ }\href@noop
  {} {\bibfield  {journal} {\bibinfo  {journal} {Phys. Rev. Lett.}\ }\textbf
  {\bibinfo {volume} {116}},\ \bibinfo {pages} {024302} (\bibinfo {year}
  {2016})}\BibitemShut {NoStop}%
\bibitem [{\citenamefont {Grossmann}\ and\ \citenamefont
  {Lohse}(2001)}]{Grossmann2001}%
  \BibitemOpen
  \bibfield  {author} {\bibinfo {author} {\bibfnamefont {S.}~\bibnamefont
  {Grossmann}}\ and\ \bibinfo {author} {\bibfnamefont {D.}~\bibnamefont
  {Lohse}},\ }\bibfield  {title} {\enquote {\bibinfo {title} {Thermal
  convection for large {P}randtl numbers},}\ }\href@noop {} {\bibfield
  {journal} {\bibinfo  {journal} {Phys. Rev. Lett.}\ }\textbf {\bibinfo
  {volume} {86}},\ \bibinfo {pages} {3316--3319} (\bibinfo {year}
  {2001})}\BibitemShut {NoStop}%
\bibitem [{\citenamefont {Stevens}\ \emph {et~al.}(2013)\citenamefont
  {Stevens}, \citenamefont {van~der Poel}, \citenamefont {Grossmann},\ and\
  \citenamefont {Lohse}}]{Stevens2013}%
  \BibitemOpen
  \bibfield  {author} {\bibinfo {author} {\bibfnamefont {R.~J. A.~M.}\
  \bibnamefont {Stevens}}, \bibinfo {author} {\bibfnamefont {E.~P.}\
  \bibnamefont {van~der Poel}}, \bibinfo {author} {\bibfnamefont
  {S.}~\bibnamefont {Grossmann}}, \ and\ \bibinfo {author} {\bibfnamefont
  {D.}~\bibnamefont {Lohse}},\ }\bibfield  {title} {\enquote {\bibinfo {title}
  {{The unifying theory of scaling in thermal convection: The updated
  prefactors}},}\ }\href@noop {} {\bibfield  {journal} {\bibinfo  {journal} {J.
  Fluid Mech.}\ }\textbf {\bibinfo {volume} {730}},\ \bibinfo {pages}
  {295--308} (\bibinfo {year} {2013})}\BibitemShut {NoStop}%
\bibitem [{\citenamefont {Grossmann}\ and\ \citenamefont
  {Lohse}(2011)}]{Grossmann2011}%
  \BibitemOpen
  \bibfield  {author} {\bibinfo {author} {\bibfnamefont {S.}~\bibnamefont
  {Grossmann}}\ and\ \bibinfo {author} {\bibfnamefont {D.}~\bibnamefont
  {Lohse}},\ }\bibfield  {title} {\enquote {\bibinfo {title} {Multiple scaling
  in the ultimate regime of thermal convection},}\ }\href@noop {} {\bibfield
  {journal} {\bibinfo  {journal} {Phys. Fluids}\ }\textbf {\bibinfo {volume}
  {23}},\ \bibinfo {pages} {045108} (\bibinfo {year} {2011})}\BibitemShut
  {NoStop}%
\bibitem [{\citenamefont {Hassanzadeh}\ \emph {et~al.}(2014)\citenamefont
  {Hassanzadeh}, \citenamefont {Chini},\ and\ \citenamefont
  {Doering}}]{Hassanzadeh2014}%
  \BibitemOpen
  \bibfield  {author} {\bibinfo {author} {\bibfnamefont {P.}~\bibnamefont
  {Hassanzadeh}}, \bibinfo {author} {\bibfnamefont {G.~P.}\ \bibnamefont
  {Chini}}, \ and\ \bibinfo {author} {\bibfnamefont {C.~R.}\ \bibnamefont
  {Doering}},\ }\bibfield  {title} {\enquote {\bibinfo {title} {Wall to wall
  optimal transport},}\ }\href@noop {} {\bibfield  {journal} {\bibinfo
  {journal} {J. Fluid Mech.}\ }\textbf {\bibinfo {volume} {751}},\ \bibinfo
  {pages} {627--662} (\bibinfo {year} {2014})}\BibitemShut {NoStop}%
\bibitem [{\citenamefont {Gibert}\ \emph {et~al.}(2006)\citenamefont {Gibert},
  \citenamefont {Pabiou}, \citenamefont {Chill\`a},\ and\ \citenamefont
  {Castaing}}]{Gibert2006}%
  \BibitemOpen
  \bibfield  {author} {\bibinfo {author} {\bibfnamefont {M.}~\bibnamefont
  {Gibert}}, \bibinfo {author} {\bibfnamefont {H.}~\bibnamefont {Pabiou}},
  \bibinfo {author} {\bibfnamefont {F.}~\bibnamefont {Chill\`a}}, \ and\
  \bibinfo {author} {\bibfnamefont {B.}~\bibnamefont {Castaing}},\ }\bibfield
  {title} {\enquote {\bibinfo {title} {High-rayleigh-number convection in a
  vertical channel},}\ }\href@noop {} {\bibfield  {journal} {\bibinfo
  {journal} {Phys. Rev. Lett.}\ }\textbf {\bibinfo {volume} {96}},\ \bibinfo
  {pages} {084501} (\bibinfo {year} {2006})}\BibitemShut {NoStop}%
\bibitem [{\citenamefont {Daya}\ and\ \citenamefont {Ecke}(2001)}]{Daya2001}%
  \BibitemOpen
  \bibfield  {author} {\bibinfo {author} {\bibfnamefont {Z.~A.}\ \bibnamefont
  {Daya}}\ and\ \bibinfo {author} {\bibfnamefont {R.~E.}\ \bibnamefont
  {Ecke}},\ }\bibfield  {title} {\enquote {\bibinfo {title} {Does turbulent
  convection feel the shape of the container?}}\ }\href@noop {} {\bibfield
  {journal} {\bibinfo  {journal} {Phys. Rev. Lett.}\ }\textbf {\bibinfo
  {volume} {87}},\ \bibinfo {pages} {184501} (\bibinfo {year}
  {2001})}\BibitemShut {NoStop}%
\bibitem [{\citenamefont {He}\ \emph {et~al.}(2011)\citenamefont {He},
  \citenamefont {Ching},\ and\ \citenamefont {Tong}}]{He2011}%
  \BibitemOpen
  \bibfield  {author} {\bibinfo {author} {\bibfnamefont {X.}~\bibnamefont
  {He}}, \bibinfo {author} {\bibfnamefont {E.~S.~C.}\ \bibnamefont {Ching}}, \
  and\ \bibinfo {author} {\bibfnamefont {P.}~\bibnamefont {Tong}},\ }\bibfield
  {title} {\enquote {\bibinfo {title} {{Locally averaged thermal dissipation
  rate in turbulent thermal convection: A decomposition into contributions from
  different temperature gradient components}},}\ }\href@noop {} {\bibfield
  {journal} {\bibinfo  {journal} {Phys. Fluids}\ }\textbf {\bibinfo {volume}
  {23}},\ \bibinfo {pages} {025106} (\bibinfo {year} {2011})}\BibitemShut
  {NoStop}%
\bibitem [{\citenamefont {Boffetta}\ and\ \citenamefont
  {Ecke}(2012)}]{Boffetta2012}%
  \BibitemOpen
  \bibfield  {author} {\bibinfo {author} {\bibfnamefont {G.}~\bibnamefont
  {Boffetta}}\ and\ \bibinfo {author} {\bibfnamefont {R.~E.}\ \bibnamefont
  {Ecke}},\ }\bibfield  {title} {\enquote {\bibinfo {title} {Two-dimensional
  turbulence},}\ }\href@noop {} {\bibfield  {journal} {\bibinfo  {journal}
  {Ann. Rev. Fluid Mech.}\ }\textbf {\bibinfo {volume} {44}},\ \bibinfo {pages}
  {427--451} (\bibinfo {year} {2012})}\BibitemShut {NoStop}%
\bibitem [{\citenamefont {Doering}\ \emph {et~al.}(2006)\citenamefont
  {Doering}, \citenamefont {Otto},\ and\ \citenamefont
  {Reznikoff}}]{Doering2006}%
  \BibitemOpen
  \bibfield  {author} {\bibinfo {author} {\bibfnamefont {C.~R.}\ \bibnamefont
  {Doering}}, \bibinfo {author} {\bibfnamefont {F.}~\bibnamefont {Otto}}, \
  and\ \bibinfo {author} {\bibfnamefont {M.~G.}\ \bibnamefont {Reznikoff}},\
  }\bibfield  {title} {\enquote {\bibinfo {title} {{Bounds on vertical heat
  transport for infinite-Prandtl-number {R}ayleigh--{B}\'{e}nard
  convection}},}\ }\href@noop {} {\bibfield  {journal} {\bibinfo  {journal} {J.
  Fluid Mech.}\ }\textbf {\bibinfo {volume} {560}},\ \bibinfo {pages}
  {229--242} (\bibinfo {year} {2006})}\BibitemShut {NoStop}%
\bibitem [{\citenamefont {Shishkina}\ and\ \citenamefont
  {Horn}(2016)}]{Shishkina2016b}%
  \BibitemOpen
  \bibfield  {author} {\bibinfo {author} {\bibfnamefont {O.}~\bibnamefont
  {Shishkina}}\ and\ \bibinfo {author} {\bibfnamefont {S.}~\bibnamefont
  {Horn}},\ }\bibfield  {title} {\enquote {\bibinfo {title} {{Thermal
  convection in inclined cylindrical containers}},}\ }\href@noop {} {\bibfield
  {journal} {\bibinfo  {journal} {J. Fluid Mech.}\ }\textbf {\bibinfo {volume}
  {790}},\ \bibinfo {pages} {R3} (\bibinfo {year} {2016})}\BibitemShut
  {NoStop}%
\bibitem [{\citenamefont {Shishkina}(2016)}]{Shishkina2016c}%
  \BibitemOpen
  \bibfield  {author} {\bibinfo {author} {\bibfnamefont {O.}~\bibnamefont
  {Shishkina}},\ }\bibfield  {title} {\enquote {\bibinfo {title} {Momentum and
  heat transport scalings in laminar vertical convection},}\ }\href@noop {}
  {\bibfield  {journal} {\bibinfo  {journal} {Phys. Rev. E}\ }\textbf {\bibinfo
  {volume} {93}},\ \bibinfo {pages} {051102(R)} (\bibinfo {year}
  {2016})}\BibitemShut {NoStop}%
\bibitem [{\citenamefont {Z\"urner}\ \emph {et~al.}(2016)\citenamefont
  {Z\"urner}, \citenamefont {Liu}, \citenamefont {Krasnov},\ and\ \citenamefont
  {Schumacher}}]{Zuerner2016}%
  \BibitemOpen
  \bibfield  {author} {\bibinfo {author} {\bibfnamefont {T.}~\bibnamefont
  {Z\"urner}}, \bibinfo {author} {\bibfnamefont {W.}~\bibnamefont {Liu}},
  \bibinfo {author} {\bibfnamefont {D.}~\bibnamefont {Krasnov}}, \ and\
  \bibinfo {author} {\bibfnamefont {J.}~\bibnamefont {Schumacher}},\ }\bibfield
   {title} {\enquote {\bibinfo {title} {Heat and momentum transfer for
  magnetoconvection in a vertical external magnetic field},}\ }\href@noop {}
  {\bibfield  {journal} {\bibinfo  {journal} {Phys. Rev. E}\ }\textbf {\bibinfo
  {volume} {94}},\ \bibinfo {pages} {043108} (\bibinfo {year}
  {2016})}\BibitemShut {NoStop}%
\bibitem [{\citenamefont {{Ahlers}}\ \emph {et~al.}(2014)\citenamefont
  {{Ahlers}}, \citenamefont {{Bodenschatz}},\ and\ \citenamefont
  {{He}}}]{Ahlers2014}%
  \BibitemOpen
  \bibfield  {author} {\bibinfo {author} {\bibfnamefont {G.}~\bibnamefont
  {{Ahlers}}}, \bibinfo {author} {\bibfnamefont {E.}~\bibnamefont
  {{Bodenschatz}}}, \ and\ \bibinfo {author} {\bibfnamefont {X.}~\bibnamefont
  {{He}}},\ }\bibfield  {title} {\enquote {\bibinfo {title} {{Logarithmic
  temperature profiles of turbulent Rayleigh--B\'enard convection in the
  classical and ultimate state for a Prandtl number of 0.8}},}\ }\href@noop {}
  {\bibfield  {journal} {\bibinfo  {journal} {J. Fluid Mech.}\ }\textbf
  {\bibinfo {volume} {758}},\ \bibinfo {pages} {436--467} (\bibinfo {year}
  {2014})}\BibitemShut {NoStop}%
\bibitem [{\citenamefont {Grossmann}\ and\ \citenamefont
  {Lohse}(2012)}]{Grossmann2012}%
  \BibitemOpen
  \bibfield  {author} {\bibinfo {author} {\bibfnamefont {S.}~\bibnamefont
  {Grossmann}}\ and\ \bibinfo {author} {\bibfnamefont {D.}~\bibnamefont
  {Lohse}},\ }\bibfield  {title} {\enquote {\bibinfo {title} {Logarithmic
  temperature profiles in the ultimate regime of thermal convection},}\
  }\href@noop {} {\bibfield  {journal} {\bibinfo  {journal} {Phys. Fluids}\
  }\textbf {\bibinfo {volume} {24}},\ \bibinfo {pages} {125103} (\bibinfo
  {year} {2012})}\BibitemShut {NoStop}%
\bibitem [{\citenamefont {Prandtl}(1905)}]{Prandtl1905}%
  \BibitemOpen
  \bibfield  {author} {\bibinfo {author} {\bibfnamefont {L.}~\bibnamefont
  {Prandtl}},\ }\bibfield  {title} {\enquote {\bibinfo {title} {{\"U}ber
  {Fl\"ussigkeitsbewegung} bei sehr kleiner {Reibung}},}\ }in\ \href@noop {}
  {\emph {\bibinfo {booktitle} {Verhandlungen des III. Int. Math. Kongr.,
  Heidelberg, 1904}}}\ (\bibinfo  {publisher} {Teubner},\ \bibinfo {year}
  {1905})\ pp.\ \bibinfo {pages} {484--491}\BibitemShut {NoStop}%
\bibitem [{\citenamefont {Landau}\ and\ \citenamefont
  {Lifshitz}(1987)}]{Landau1987}%
  \BibitemOpen
  \bibfield  {author} {\bibinfo {author} {\bibfnamefont {L.~D.}\ \bibnamefont
  {Landau}}\ and\ \bibinfo {author} {\bibfnamefont {E.~M.}\ \bibnamefont
  {Lifshitz}},\ }\href@noop {} {\emph {\bibinfo {title} {Fluid Mechanics}}},\
  \bibinfo {edition} {2nd}\ ed.,\ \bibinfo {series} {Course of Theoretical
  Physics}, Vol.~\bibinfo {volume} {6}\ (\bibinfo  {publisher} {Butterworth
  Heinemann},\ \bibinfo {year} {1987})\BibitemShut {NoStop}%
\bibitem [{\citenamefont {Shishkina}\ and\ \citenamefont
  {Thess}(2009)}]{Shishkina2009}%
  \BibitemOpen
  \bibfield  {author} {\bibinfo {author} {\bibfnamefont {O.}~\bibnamefont
  {Shishkina}}\ and\ \bibinfo {author} {\bibfnamefont {A.}~\bibnamefont
  {Thess}},\ }\bibfield  {title} {\enquote {\bibinfo {title} {Mean temperature
  profiles in turbulent {R}ayleigh--{B}\'{e}nard convection of water},}\
  }\href@noop {} {\bibfield  {journal} {\bibinfo  {journal} {J. Fluid Mech.}\
  }\textbf {\bibinfo {volume} {663}},\ \bibinfo {pages} {449--460} (\bibinfo
  {year} {2009})}\BibitemShut {NoStop}%
\bibitem [{\citenamefont {Shi}\ \emph {et~al.}(2012)\citenamefont {Shi},
  \citenamefont {Emran},\ and\ \citenamefont {Schumacher}}]{Shi2012}%
  \BibitemOpen
  \bibfield  {author} {\bibinfo {author} {\bibfnamefont {N.}~\bibnamefont
  {Shi}}, \bibinfo {author} {\bibfnamefont {M.~S.}\ \bibnamefont {Emran}}, \
  and\ \bibinfo {author} {\bibfnamefont {J}~\bibnamefont {Schumacher}},\
  }\bibfield  {title} {\enquote {\bibinfo {title} {Boundary layer structure in
  turbulent {R}ayleigh--{B}\'{e}nard convection},}\ }\href@noop {} {\bibfield
  {journal} {\bibinfo  {journal} {J. Fluid Mech.}\ }\textbf {\bibinfo {volume}
  {706}},\ \bibinfo {pages} {5--33} (\bibinfo {year} {2012})}\BibitemShut
  {NoStop}%
\bibitem [{\citenamefont {Scheel}\ \emph {et~al.}(2012)\citenamefont {Scheel},
  \citenamefont {Kim},\ and\ \citenamefont {White}}]{Scheel2012}%
  \BibitemOpen
  \bibfield  {author} {\bibinfo {author} {\bibfnamefont {J.~D.}\ \bibnamefont
  {Scheel}}, \bibinfo {author} {\bibfnamefont {E.}~\bibnamefont {Kim}}, \ and\
  \bibinfo {author} {\bibfnamefont {K.~R.}\ \bibnamefont {White}},\ }\bibfield
  {title} {\enquote {\bibinfo {title} {Thermal and viscous boundary layers in
  turbulent {R}ayleigh--{B}\'{e}nard convection},}\ }\href@noop {} {\bibfield
  {journal} {\bibinfo  {journal} {J. Fluid Mech.}\ }\textbf {\bibinfo {volume}
  {711}},\ \bibinfo {pages} {281--305} (\bibinfo {year} {2012})}\BibitemShut
  {NoStop}%
\bibitem [{\citenamefont {Stevens}\ \emph {et~al.}(2012)\citenamefont
  {Stevens}, \citenamefont {Zhou}, \citenamefont {Grossmann}, \citenamefont
  {Verzicco}, \citenamefont {Xia},\ and\ \citenamefont {Lohse}}]{Stevens2012}%
  \BibitemOpen
  \bibfield  {author} {\bibinfo {author} {\bibfnamefont {R.~J. A.~M.}\
  \bibnamefont {Stevens}}, \bibinfo {author} {\bibfnamefont {Q.}~\bibnamefont
  {Zhou}}, \bibinfo {author} {\bibfnamefont {S.}~\bibnamefont {Grossmann}},
  \bibinfo {author} {\bibfnamefont {R.}~\bibnamefont {Verzicco}}, \bibinfo
  {author} {\bibfnamefont {K.-Q.}\ \bibnamefont {Xia}}, \ and\ \bibinfo
  {author} {\bibfnamefont {D.}~\bibnamefont {Lohse}},\ }\bibfield  {title}
  {\enquote {\bibinfo {title} {{Thermal boundary layer profiles in turbulent
  Rayleigh--B\'enard convection in a cylindrical sample}},}\ }\href@noop {}
  {\bibfield  {journal} {\bibinfo  {journal} {Phys. Rev. E}\ }\textbf {\bibinfo
  {volume} {85}},\ \bibinfo {pages} {027301} (\bibinfo {year}
  {2012})}\BibitemShut {NoStop}%
\bibitem [{\citenamefont {Kaczorowski}\ \emph {et~al.}(2011)\citenamefont
  {Kaczorowski}, \citenamefont {Shishkina}, \citenamefont {Shishkin},
  \citenamefont {Wagner},\ and\ \citenamefont {Xia}}]{Kaczorowski2011}%
  \BibitemOpen
  \bibfield  {author} {\bibinfo {author} {\bibfnamefont {M.}~\bibnamefont
  {Kaczorowski}}, \bibinfo {author} {\bibfnamefont {O.}~\bibnamefont
  {Shishkina}}, \bibinfo {author} {\bibfnamefont {A.}~\bibnamefont {Shishkin}},
  \bibinfo {author} {\bibfnamefont {C.}~\bibnamefont {Wagner}}, \ and\ \bibinfo
  {author} {\bibfnamefont {K.-Q.}\ \bibnamefont {Xia}},\ }\bibfield  {title}
  {\enquote {\bibinfo {title} {{Analysis of the large-scale circulation and the
  boundary layers in turbulent Rayleigh--B\'enard convection}},}\ }in\
  \href@noop {} {\emph {\bibinfo {booktitle} {Direct and Large-Eddy Simulation
  VIII}}},\ \bibinfo {editor} {edited by\ \bibinfo {editor} {\bibfnamefont
  {H.}~\bibnamefont {Kuerten}}, \bibinfo {editor} {\bibfnamefont
  {B.}~\bibnamefont {Geurts}}, \bibinfo {editor} {\bibfnamefont
  {V.}~\bibnamefont {Armenio}}, \ and\ \bibinfo {editor} {\bibfnamefont
  {J.}~\bibnamefont {Fr\"ohlich}}}\ (\bibinfo  {publisher} {Springer},\
  \bibinfo {year} {2011})\ pp.\ \bibinfo {pages} {383--388}\BibitemShut
  {NoStop}%
\bibitem [{\citenamefont {Ovsyannikov}\ \emph {et~al.}(2016)\citenamefont
  {Ovsyannikov}, \citenamefont {Krasnov}, \citenamefont {Emran},\ and\
  \citenamefont {Schumacher}}]{Ovsyannikov2016}%
  \BibitemOpen
  \bibfield  {author} {\bibinfo {author} {\bibfnamefont {M.}~\bibnamefont
  {Ovsyannikov}}, \bibinfo {author} {\bibfnamefont {D.}~\bibnamefont
  {Krasnov}}, \bibinfo {author} {\bibfnamefont {M.~S.}\ \bibnamefont {Emran}},
  \ and\ \bibinfo {author} {\bibfnamefont {J.}~\bibnamefont {Schumacher}},\
  }\bibfield  {title} {\enquote {\bibinfo {title} {{Combined effects of
  prescribed pressure gradient and buoyancy in boundary layer of turbulent
  Rayleigh--B\'enard convection}},}\ }\href@noop {} {\bibfield  {journal}
  {\bibinfo  {journal} {Eur. J. Mech. (B/Fluids)}\ }\textbf {\bibinfo {volume}
  {57}},\ \bibinfo {pages} {64--74} (\bibinfo {year} {2016})}\BibitemShut
  {NoStop}%
\bibitem [{\citenamefont {Zhou}\ and\ \citenamefont {Xia}(2010)}]{Zhou2010}%
  \BibitemOpen
  \bibfield  {author} {\bibinfo {author} {\bibfnamefont {Q.}~\bibnamefont
  {Zhou}}\ and\ \bibinfo {author} {\bibfnamefont {K.-Q.}\ \bibnamefont {Xia}},\
  }\bibfield  {title} {\enquote {\bibinfo {title} {Measured instantaneous
  viscous boundary layer in turbulent {R}ayleigh--{B}\'{e}nard convection},}\
  }\href@noop {} {\bibfield  {journal} {\bibinfo  {journal} {Phys. Rev. Lett.}\
  }\textbf {\bibinfo {volume} {104}},\ \bibinfo {pages} {104301} (\bibinfo
  {year} {2010})}\BibitemShut {NoStop}%
\bibitem [{\citenamefont {Shishkina}\ \emph {et~al.}(2013)\citenamefont
  {Shishkina}, \citenamefont {Horn},\ and\ \citenamefont
  {Wagner}}]{Shishkina2013}%
  \BibitemOpen
  \bibfield  {author} {\bibinfo {author} {\bibfnamefont {O.}~\bibnamefont
  {Shishkina}}, \bibinfo {author} {\bibfnamefont {S.}~\bibnamefont {Horn}}, \
  and\ \bibinfo {author} {\bibfnamefont {S.}~\bibnamefont {Wagner}},\
  }\bibfield  {title} {\enquote {\bibinfo {title} {{Falkner-Skan boundary layer
  approximation in Rayleigh--B\'enard convection}},}\ }\href@noop {} {\bibfield
   {journal} {\bibinfo  {journal} {J. Fluid Mech.}\ }\textbf {\bibinfo {volume}
  {730}},\ \bibinfo {pages} {442--463} (\bibinfo {year} {2013})}\BibitemShut
  {NoStop}%
\bibitem [{\citenamefont {Falkner}\ and\ \citenamefont
  {Skan}(1931)}]{Falkner1931}%
  \BibitemOpen
  \bibfield  {author} {\bibinfo {author} {\bibfnamefont {V.~M.}\ \bibnamefont
  {Falkner}}\ and\ \bibinfo {author} {\bibfnamefont {S.~W.}\ \bibnamefont
  {Skan}},\ }\bibfield  {title} {\enquote {\bibinfo {title} {Some approximate
  solutions of the boundary layer equations},}\ }\href@noop {} {\bibfield
  {journal} {\bibinfo  {journal} {Phil. Mag.}\ }\textbf {\bibinfo {volume}
  {12}},\ \bibinfo {pages} {865--896} (\bibinfo {year} {1931})}\BibitemShut
  {NoStop}%
\bibitem [{\citenamefont {Shishkina}\ \emph {et~al.}(2014)\citenamefont
  {Shishkina}, \citenamefont {Wagner},\ and\ \citenamefont
  {Horn}}]{Shishkina2014}%
  \BibitemOpen
  \bibfield  {author} {\bibinfo {author} {\bibfnamefont {O.}~\bibnamefont
  {Shishkina}}, \bibinfo {author} {\bibfnamefont {S.}~\bibnamefont {Wagner}}, \
  and\ \bibinfo {author} {\bibfnamefont {S.}~\bibnamefont {Horn}},\ }\bibfield
  {title} {\enquote {\bibinfo {title} {Influence of the angle between the wind
  and the isothermal surfaces on the boundary layer structures in turbulent
  thermal convection},}\ }\href@noop {} {\bibfield  {journal} {\bibinfo
  {journal} {Phys. Rev. E}\ }\textbf {\bibinfo {volume} {89}},\ \bibinfo
  {pages} {033014} (\bibinfo {year} {2014})}\BibitemShut {NoStop}%
\bibitem [{\citenamefont {Shraiman}\ and\ \citenamefont
  {Siggia}(1990)}]{Shraiman1990}%
  \BibitemOpen
  \bibfield  {author} {\bibinfo {author} {\bibfnamefont {B.~I.}\ \bibnamefont
  {Shraiman}}\ and\ \bibinfo {author} {\bibfnamefont {E.~D.}\ \bibnamefont
  {Siggia}},\ }\bibfield  {title} {\enquote {\bibinfo {title} {Heat transport
  in high-{R}ayleigh-number convection},}\ }\href@noop {} {\bibfield  {journal}
  {\bibinfo  {journal} {Phys. Rev. A}\ }\textbf {\bibinfo {volume} {42}},\
  \bibinfo {pages} {3650--3653} (\bibinfo {year} {1990})}\BibitemShut {NoStop}%
\bibitem [{\citenamefont {Ching}(1997)}]{Ching1997}%
  \BibitemOpen
  \bibfield  {author} {\bibinfo {author} {\bibfnamefont {E.~S.~C.}\
  \bibnamefont {Ching}},\ }\bibfield  {title} {\enquote {\bibinfo {title}
  {{Heat flux and shear rate in turbulent convection}},}\ }\href@noop {}
  {\bibfield  {journal} {\bibinfo  {journal} {Phys. Rev. E}\ }\textbf {\bibinfo
  {volume} {55}},\ \bibinfo {pages} {1189--1192} (\bibinfo {year}
  {1997})}\BibitemShut {NoStop}%
\bibitem [{\citenamefont {Shishkina}\ \emph {et~al.}(2015)\citenamefont
  {Shishkina}, \citenamefont {Horn}, \citenamefont {Wagner},\ and\
  \citenamefont {Ching}}]{Shishkina2015}%
  \BibitemOpen
  \bibfield  {author} {\bibinfo {author} {\bibfnamefont {O.}~\bibnamefont
  {Shishkina}}, \bibinfo {author} {\bibfnamefont {S.}~\bibnamefont {Horn}},
  \bibinfo {author} {\bibfnamefont {S.}~\bibnamefont {Wagner}}, \ and\ \bibinfo
  {author} {\bibfnamefont {E.~S.~C.}\ \bibnamefont {Ching}},\ }\bibfield
  {title} {\enquote {\bibinfo {title} {{Thermal boundary layer equation for
  turbulent Rayleigh--B\'enard convection}},}\ }\href@noop {} {\bibfield
  {journal} {\bibinfo  {journal} {Phys. Rev. Lett.}\ }\textbf {\bibinfo
  {volume} {114}},\ \bibinfo {pages} {114302} (\bibinfo {year}
  {2015})}\BibitemShut {NoStop}%
\bibitem [{\citenamefont {Antonia}\ and\ \citenamefont
  {Kim}(1991)}]{Antonia1991}%
  \BibitemOpen
  \bibfield  {author} {\bibinfo {author} {\bibfnamefont {R.~A.}\ \bibnamefont
  {Antonia}}\ and\ \bibinfo {author} {\bibfnamefont {J.}~\bibnamefont {Kim}},\
  }\bibfield  {title} {\enquote {\bibinfo {title} {{Turbulent Prandtl number in
  the near-wall region of a turbulent channel flow}},}\ }\href@noop {}
  {\bibfield  {journal} {\bibinfo  {journal} {Int. J. Heat Mass Transfer}\
  }\textbf {\bibinfo {volume} {34}},\ \bibinfo {pages} {1905--1908} (\bibinfo
  {year} {1991})}\BibitemShut {NoStop}%
\bibitem [{\citenamefont {Ahlers}\ \emph {et~al.}(2012)\citenamefont {Ahlers},
  \citenamefont {Bodenschatz}, \citenamefont {Funfschilling}, \citenamefont
  {Grossmann}, \citenamefont {He}, \citenamefont {Lohse}, \citenamefont
  {Stevens},\ and\ \citenamefont {Verzicco}}]{Ahlers2012}%
  \BibitemOpen
  \bibfield  {author} {\bibinfo {author} {\bibfnamefont {G.}~\bibnamefont
  {Ahlers}}, \bibinfo {author} {\bibfnamefont {E.}~\bibnamefont {Bodenschatz}},
  \bibinfo {author} {\bibfnamefont {D.}~\bibnamefont {Funfschilling}}, \bibinfo
  {author} {\bibfnamefont {S.}~\bibnamefont {Grossmann}}, \bibinfo {author}
  {\bibfnamefont {X.}~\bibnamefont {He}}, \bibinfo {author} {\bibfnamefont
  {D.}~\bibnamefont {Lohse}}, \bibinfo {author} {\bibfnamefont {R.J.A.M.}\
  \bibnamefont {Stevens}}, \ and\ \bibinfo {author} {\bibfnamefont
  {R.}~\bibnamefont {Verzicco}},\ }\bibfield  {title} {\enquote {\bibinfo
  {title} {{Logarithmic temperature profiles in turbulent Rayleigh--B\'enard
  convection}},}\ }\href@noop {} {\bibfield  {journal} {\bibinfo  {journal}
  {Phys. Rev. Lett.}\ }\textbf {\bibinfo {volume} {109}},\ \bibinfo {pages}
  {114501} (\bibinfo {year} {2012})}\BibitemShut {NoStop}%
\bibitem [{\citenamefont {Ahlers}\ \emph {et~al.}(2014)\citenamefont {Ahlers},
  \citenamefont {Bodenschatz},\ and\ \citenamefont {He}}]{He2014}%
  \BibitemOpen
  \bibfield  {author} {\bibinfo {author} {\bibfnamefont {G.}~\bibnamefont
  {Ahlers}}, \bibinfo {author} {\bibfnamefont {E.}~\bibnamefont {Bodenschatz}},
  \ and\ \bibinfo {author} {\bibfnamefont {X.}~\bibnamefont {He}},\ }\bibfield
  {title} {\enquote {\bibinfo {title} {Logarithmic temperature profiles of
  turbulent {R}ayleigh-{B}\'{e}nard convection in the classical and ultimate
  state for a prandtl number of 0.8},}\ }\href@noop {} {\bibfield  {journal}
  {\bibinfo  {journal} {J. Fluid Mech.}\ }\textbf {\bibinfo {volume} {758}},\
  \bibinfo {pages} {436--467} (\bibinfo {year} {2014})}\BibitemShut {NoStop}%
\bibitem [{\citenamefont {Prandtl}(1925)}]{Prandtl1925}%
  \BibitemOpen
  \bibfield  {author} {\bibinfo {author} {\bibfnamefont {L.}~\bibnamefont
  {Prandtl}},\ }\bibfield  {title} {\enquote {\bibinfo {title} {Bericht
  {\"u}ber {{Untersuchungen}} zur ausgebildeten {{Turbulenz}}},}\ }\href@noop
  {} {\bibfield  {journal} {\bibinfo  {journal} {Z. Angew. Math. Mech.}\
  }\textbf {\bibinfo {volume} {5}},\ \bibinfo {pages} {136--139} (\bibinfo
  {year} {1925})}\BibitemShut {NoStop}%
\bibitem [{\citenamefont {Ching}\ \emph {et~al.}(2017)\citenamefont {Ching},
  \citenamefont {Dung},\ and\ \citenamefont {Shishkina}}]{Ching2017}%
  \BibitemOpen
  \bibfield  {author} {\bibinfo {author} {\bibfnamefont {E.~S.~C.}\
  \bibnamefont {Ching}}, \bibinfo {author} {\bibfnamefont {O.-Y.}\ \bibnamefont
  {Dung}}, \ and\ \bibinfo {author} {\bibfnamefont {O.}~\bibnamefont
  {Shishkina}},\ }\bibfield  {title} {\enquote {\bibinfo {title} {{Fluctuating
  thermal boundary layers and heat transfer in turbulent Rayleigh--B\'enard
  convection}},}\ }\href@noop {} {\bibfield  {journal} {\bibinfo  {journal} {J.
  Stat. Phys.}\ }\textbf {\bibinfo {volume} {167}},\ \bibinfo {pages}
  {626--635} (\bibinfo {year} {2017})}\BibitemShut {NoStop}%
\bibitem [{\citenamefont {Kooij}\ \emph {et~al.}(2017)\citenamefont {Kooij},
  \citenamefont {Botchev}, \citenamefont {Frederix}, \citenamefont {Geurts},
  \citenamefont {Horn}, \citenamefont {Lohse}, \citenamefont {van~der Poel},
  \citenamefont {Shishkina}, \citenamefont {Stevens},\ and\ \citenamefont
  {Verzicco}}]{Kooij2017}%
  \BibitemOpen
  \bibfield  {author} {\bibinfo {author} {\bibfnamefont {G.~L.}\ \bibnamefont
  {Kooij}}, \bibinfo {author} {\bibfnamefont {M.~A.}\ \bibnamefont {Botchev}},
  \bibinfo {author} {\bibfnamefont {E.~M.A.}\ \bibnamefont {Frederix}},
  \bibinfo {author} {\bibfnamefont {B.~J.}\ \bibnamefont {Geurts}}, \bibinfo
  {author} {\bibfnamefont {S.}~\bibnamefont {Horn}}, \bibinfo {author}
  {\bibfnamefont {D.}~\bibnamefont {Lohse}}, \bibinfo {author} {\bibfnamefont
  {E.~P.}\ \bibnamefont {van~der Poel}}, \bibinfo {author} {\bibfnamefont
  {O.}~\bibnamefont {Shishkina}}, \bibinfo {author} {\bibfnamefont {R.~J.
  A.~M.}\ \bibnamefont {Stevens}}, \ and\ \bibinfo {author} {\bibfnamefont
  {R.}~\bibnamefont {Verzicco}},\ }\bibfield  {title} {\enquote {\bibinfo
  {title} {{Comparison of computational codes for direct numerical simulations
  of turbulent Rayleigh--B\'enard convection}},}\ }\href@noop {} {\bibfield
  {journal} {\bibinfo  {journal} {Comp. Fluids, submitted}\ } (\bibinfo {year}
  {2017})}\BibitemShut {NoStop}%
\bibitem [{\citenamefont {Shishkina}\ \emph {et~al.}(2010)\citenamefont
  {Shishkina}, \citenamefont {Stevens}, \citenamefont {Grossmann},\ and\
  \citenamefont {Lohse}}]{Shishkina2010}%
  \BibitemOpen
  \bibfield  {author} {\bibinfo {author} {\bibfnamefont {O.}~\bibnamefont
  {Shishkina}}, \bibinfo {author} {\bibfnamefont {R.~J. A.~M.}\ \bibnamefont
  {Stevens}}, \bibinfo {author} {\bibfnamefont {S.}~\bibnamefont {Grossmann}},
  \ and\ \bibinfo {author} {\bibfnamefont {D.}~\bibnamefont {Lohse}},\
  }\bibfield  {title} {\enquote {\bibinfo {title} {Boundary layer structure in
  turbulent thermal convection and its consequences for the required numerical
  resolution},}\ }\href@noop {} {\bibfield  {journal} {\bibinfo  {journal} {New
  J. Phys.}\ }\textbf {\bibinfo {volume} {12}},\ \bibinfo {pages} {075022}
  (\bibinfo {year} {2010})}\BibitemShut {NoStop}%
\bibitem [{\citenamefont {Scheel}\ and\ \citenamefont
  {Schumacher}(2016)}]{Scheel2016}%
  \BibitemOpen
  \bibfield  {author} {\bibinfo {author} {\bibfnamefont {J.~D.}\ \bibnamefont
  {Scheel}}\ and\ \bibinfo {author} {\bibfnamefont {J.}~\bibnamefont
  {Schumacher}},\ }\bibfield  {title} {\enquote {\bibinfo {title} {{Global and
  local statistics in turbulent convection at low Prandtl numbers}},}\
  }\href@noop {} {\bibfield  {journal} {\bibinfo  {journal} {J. Fluid Mech.}\
  }\textbf {\bibinfo {volume} {802}},\ \bibinfo {pages} {147--173} (\bibinfo
  {year} {2016})}\BibitemShut {NoStop}%
\bibitem [{\citenamefont {Schumacher}\ \emph {et~al.}(2016)\citenamefont
  {Schumacher}, \citenamefont {Bandaru}, \citenamefont {Pandey},\ and\
  \citenamefont {Scheel}}]{Schumacher2016}%
  \BibitemOpen
  \bibfield  {author} {\bibinfo {author} {\bibfnamefont {J.}~\bibnamefont
  {Schumacher}}, \bibinfo {author} {\bibfnamefont {V.}~\bibnamefont {Bandaru}},
  \bibinfo {author} {\bibfnamefont {A.}~\bibnamefont {Pandey}}, \ and\ \bibinfo
  {author} {\bibfnamefont {J.~D.}\ \bibnamefont {Scheel}},\ }\bibfield  {title}
  {\enquote {\bibinfo {title} {{Transitional boundary layers in
  low-Prandtl-number convection}},}\ }\href@noop {} {\bibfield  {journal}
  {\bibinfo  {journal} {Phys. Rev. Fluids}\ }\textbf {\bibinfo {volume} {1}},\
  \bibinfo {pages} {084402} (\bibinfo {year} {2016})}\BibitemShut {NoStop}%
\end{thebibliography}

\end{document}